\documentclass[iop]{emulateapj}
\usepackage{amsmath}
\usepackage{natbib}
\usepackage{graphicx}

\begin{document}

\title{X-ray Polarization from High Mass X-ray Binaries}

\author{T. Kallman}

\affil{NASA/GSFC, Code 662, Greenbelt MD 20771}
	
\author{A. Dorodnitsyn$^4$}

\affil{Department of Astronomy, University of Maryland, College Park MD 20742}
\altaffiltext{4}{Space Research Institute, Profsoyuznaya st., 84/32, 117997, Moscow, Russia}

\author{J. Blondin}

\affil{Department of Physics, North Carolina State University, Raleigh, NC 27695-8202}

\begin{abstract}
X-ray astronomy allows study of objects which may be associated with compact objects, i.e. neutron stars 
or black holes, and also may contain strong magnetic fields.  Such objects are categorically non-spherical, 
and likely non-circular when projected on the sky.  Polarization allows study of such geometric effects, 
and X-ray polarimetry is likely to become feasible for a significant number of sources in the future.
Potential targets for future X-ray polarization observations are the high-mass X-ray binaries  (HMXBs), which consist of a compact object in orbit with an early type star.
In this paper we show that X-ray polarization from HMXBs has a distinct signature which depends on the
source inclination and orbital phase.  The presence of the X-ray source displaced from the star creates
linear polarization even if the primary wind is spherically symmetric whenever the system is viewed away
from conjunction.   Direct X-rays dilute this polarization whenever the X-ray source is not eclipsed;
at mid-eclipse the net polarization is expected to be small or zero if the wind is circularly
symmetric around the line of centers.    Resonance line scattering increases the scattering fraction,
often by large factors, over the energy band spanned by resonance lines.   Real winds are not expected
to be spherically symmetric, or circularly symmetric around the line of centers, owing to the combined
effects of the compact object gravity and ionization on the wind hydrodynamics.  A sample calculation
shows that this creates polarization fractions ranging up to tens of percent at mid-eclipse.
\end{abstract}

\section{Introduction}

High mass X-ray binaries (HMXBs) are among the brightest X-ray 
sources in the sky.  They consist of an accreting compact object (a neutron star 
or black hole) in orbit with an early type star (O or B star).
They represent an important stage in the evolution of binary stars with
early-type components, and they dominate the X-ray output of galaxies with
young stellar populations.  HMXBs emit X-rays 
via accretion onto the compact object by gas from the strong wind from the companion star, 
or from Roche lobe overflow. The X-rays interact with the accretion flow and stellar 
wind, producing observable signatures in timing and spectra.  This provides
a means to study the hot star wind and the accretion flow and the interaction
of X-rays with these structures.  One consequence of this interaction is the
production of polarized X-rays via scattering.  Polarization provides unique
information into the geometry of the stars, wind and scattering region.
In this paper, we 
explore the polarization signatures associated with this interaction.

There are $\simeq$10 HMXB sources 
with properties of their stellar winds and binary orbit which are well enough determined
for quantitative modeling of the interaction of the accretion flow with X-rays \citep{Kape98}.
The number of known HMXBs is many times greater when sources with less well constrained properties are included
\citep{Liu06,Walt15}; these include sources discovered by INTEGRAL which are seen primarily at hard X-ray energies
owing to obscuration, and transient sources which are detected primarily via long time-baseline
monitoring.    The most thoroughly studied systems  include the  objects Cyg X-1, Vela X-1, and Cen X-3. 
The majority of HMXBs contain pulsating neutron stars.  If so, orbital eclipses, X-ray pulse timing and primary radial velocities
\citep{vand07} provides strong  constraints on the orbital 
separations and masses.  Simple estimates suggest that the 
accretion rate onto the compact object can be provided by the wind from the 
companion star in fewer than half the sources,  while Roche lobe overflow is 
required in the remainder \citep{Cont78, Whit85,Kape98}.   Though they are outnumbered 
by low mass X-ray binaries in our galaxy, these sources dominate the X-ray 
output from galaxies with larger star formation rates \citep{Fabb06}.
Their mass transfer can be very rapid; in a system in which the primary has a radiative envelope
and fills its Roche lobe, the primary radius will shrink as a consequence of mass transfer,
but if the mass transfer is conservative the Roche lobe will shrink more rapidly, potentially
leading to a common envelope phase.  If the primary is an evolved star with a convective envelope
this outcome is more probable \citep{Taam00}.  Common envelope evolution when the primary is an evolved star
is likely to lead to ejection of much of the envelope of the primary, leaving a binary consisting of a
neutron star and the core of the primary.   
Whether due to nuclear evolution or to unstable mass transfer, their evolutionary lifetimes 
must be short ($\leq 10^5$ yrs).    

The wind from the companion absorbs and scatters the X-rays from the accreting 
compact object in HMXBs, and the column densities traversed by the X-rays 
range from $\sim 10^{21} - 10^{24}$cm$^{-2}$.  The X-ray source luminosities 
range from $\sim 10^{35} - 10^{38}$ erg s$^{-1}$.  In sources where the X-ray 
source is luminous and the wind is weak, the wind is ionized throughout most 
of the region between the two stars.  In systems with more massive winds 
and weaker X-ray sources X-ray-ionization effects are a perturbation 
on the wind properties.
Key questions about HMXBs include the nature of the compact object: its mass, 
variability and intrinsic radiation pattern.  Also of interest are the properties of 
the companion star wind and its uniformity in density and temperature.
The X-ray source can provide a probe for the 
study of wind regions which are only weakly affected by ionization or the 
gravity of the compact object.   


Many of the X-ray properties of HMXBs are affected by the stellar wind and accretion flow. 
This includes the variability around the orbit, which shows gradual eclipse transitions due 
to photoelectric absorption in the wind \citep{Clar88}, enhanced absorption at late 
orbital phases due to a wake or stream trailing the compact object \citep{Kall82, Wata06}.  
The wind and accretion flow also provide torque to the compact object and regulate its 
spin or angular momentum.   The structure of the wind and accretion flow is 
uncertain; winds from comparable single O or B stars have radiatively driven winds \citep{Lame99}
which are affected by instabilities \citep{Owocki88}, shocks \citep{LucyWhite80}  and clumping \citep{Cohen06};
observations provide the most reliable means for understanding these processes, via fitting 
of parameterized  models.  In addition, the compact object affects the gas flow, via its gravity
and also X-ray heating and ionization.  

The next new astrophysical window will be the advent of measurements 
of X-ray polarization \citep{Jaho14,Weis13}.  Among other things, polarization allows for potentially 
sensitive tests of the geometry of astrophysical sources, on scales which are 
far too small to be imaged directly.  
There is only one source in the sky whose X-ray polarization is known, the 
Crab nebula \citep{Weis78}.  Owing to visibility constraints, it is likely 
that the first astronomical X-ray polarimetry observations will be of objects 
which have never before been observed with this technique.   This motivates 
thorough and accurate modeling of the polarization properties of the brightest 
and (otherwise) best understood X-ray sources, for use as calibrators and test sources for X-ray polarimetry. 
HMXBs are among the sources best suited for this.  Their orbital elements are relatively well understood
and their orbital variability provides a predictably changing view with respect to an important source
of polarization: the strong stellar wind from the companion star.   
X-ray polarimetry provides a means to probe the structure and the physical processes 
occuring in and near compact objects, on length scales too small to be imaged directly.
In spite of scanty astrophysical detections so far \citep{Weis78} it is of interest to 
consider the possible signals and diagnostic use of X-ray polarization in known classes 
of cosmic sources, as it is likely that instruments with improved sensitivity
will become available eventually.

Polarization properties of HMXBs have been discussed previously in the context 
of scattering of optical and UV from the primary star \citep{Brow77, Brow78, Rudy78}. 
These authors predicted that the linear polarization has a characteristic variability around the 
binary orbit, associated with scattering in the stellar envelope which is distorted by 
the gravity of the compact object.  This variability can be used as a diagnostic of the 
shape of the star, and also of the inclination of the binary orbit.   These effects are observed 
at the predicted level in some systems; their absence in other cases suggests that the optical light 
may be affected by contributions from structures whose shape is not symmetric around the line 
of centers.  Related work has been carried out by \citet{Alma99,Igna09,Nofi14}.
Useful formalism for the calculation of polarization has been provided by \citet{Matt96}.
The polarization properties of the accretion columns in X-ray pulsars have been calculated by
\citep{Mesz88}. Predictions of X-ray polarization by the large scale structures in binaries has been carried out for
distorted stars \citep{Ange69}; for Compton scattering within the accretion column of magnetic cataclysmic variables
\citep{Mcna08}, following up on pioneering work by \cite{Matt04}.

X-rays have an important advantage over optical light which is that the source is almost certainly 
compact compared with the size of the binary system.  That is, we expect X-rays to be emitted from 
a region smaller than the Alfven shell, defined as the distance from the compact object where 
the pressure due to a dipole magnetic field balances the ram pressure of accretion.  
For fields $\sim 10^{12}$ Gauss this region has a size $\sim 10^8$ cm \citep{Lamb73}.
Far outside this region the influence of the field on the dynamics is negligible.
The Alfven shell size can be compared with the primary star radius and orbital separation
which are  $\sim 10^{12}$ cm.  This disparity in sizes greatly simplifies the geometrical 
considerations associated with the use of X-ray polarization since the X-ray source is
effectively a point source when considering the effect of the scattering by the stellar wind.
 Calculating the polarization from these structures is straightforward, 
given certain simplifying assumptions, though it is important to take into account the 
effects of atomic absorption and resonance line scattering, which in turn depends on the 
gas dynamics,  along with electron scattering.

The polarization properties of HMXBs will be affected by the intrinsic polarization 
of the compact X-ray source, and by the polarization imprinted by the stellar 
wind and accretion flow.  These can be distinguished by their differing  
variability behavior:  the compact source generally varies on a pulsation  timescale for
sources containing a pulsar, which is $\sim$ seconds -- minutes.
The very few sources which contain black holes also have their strongest intrinsic variability
on comparably short timescales.  Wind variability, on the other hand,
is associated with the orbital timescale ($\sim$ days) or possibly on the wind flow timescale which is
$\geq$hours.   In some HMXBs the X-ray source is luminous enough to ionize the wind almost completely, 
so the light  observed during and near eclipse, and its polarization, is 
dominated by electron scattering.

In this paper 
we present a general discussion of the behavior of X-ray polarization from HMXBs 
as a function of the parameters of the system and the viewing position.  
We present calculations of the polarization 
signature for various simple analytic models for the wind density and ionization structure,
and discuss the dependence of these on parameters:  wind optical depth, inclination, and
viewing direction or orbital phase.   
We also explore the effects of wind hydrodynamics, i.e. departures from spherical symmetry due
to the effects of the X-ray source gravity and heating.  We utilize sample three-dimensional
dynamical models.   We take the 
intrinsic polarization of the compact object as unpolarized and explore the combined effects of
geometry and scattering physics on the predicted linear polarization in the 0.1 - 10 keV X-ray band.

 In section \ref{sec1} we present numerical calculations 
for spherical winds using only electron scattering.  The effects of resonance line scattering
are discussed in section \ref{sec3}.
In section \ref{sec3b} we present models which include photoelectric absorption 
and an ensemble of resonance scatterers.
In section \ref{sec4} we present models utilizing three-dimensional hydrodynamic models for the 
wind density and velocity field.  We use these to derive approximate predictions for the polarization
levels expected for several well known systems in section \ref{sec5}.  The Appendix presents
simple analytic estimates of the polarization for idealized HMXB conditions;
many of these mirror earlier results.   The computational techniques and level of detail we employ 
are similar to those of our previous work \citep{DoraKall10} on the polarization from warm absorbers in 
Syefert galaxies.

\section{Spherical Winds}
\label{sec1}

\begin{figure*}[p] 
\includegraphics*[angle=0, scale=0.5]{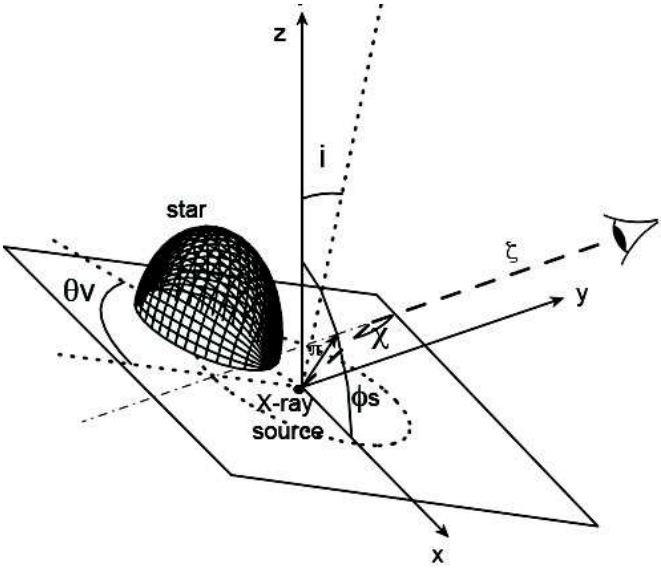}
\caption{\label{fig1} Schematic of high mass X-ray binary (HMXB) as viewed from above the orbital plane,
 showing typical orbital separation and companion 
star size. Coordinates are labeled according to their use in the Appendix.}
\end{figure*}

The basic geometry of an HMXB is illustrated in figure \ref{fig1}.
This shows  a schematic of an HMXB with a typical orbital separation and a sample 
X-ray ionized zone.  It also shows the geometry we use when calculating the
polarization:  the X-ray source is at the origin and the primary orbits in a plane
inclined by an angle $i$ to the line of sight.  The orbital phase is described by the angle $\Theta_V$ relative
to the line of centers.  The system is viewed along the y axis.

In this section we will consider the simple case in which the wind density is spherically
symmetric about the primary star, and the only interaction between the X-rays and
the wind is electron scattering.  If so, the scattering phase function follows the
Rayleigh form \citep{ChandraRadTransfer}.  The scattered emissivity at each point
depends only on the local gas density and on the flux from the X-ray source.
We assume the X-ray source radiates isotropically with a constant total luminosity
$L_0$ and is unpolarized.

We calculate the polarization signatures in the form of the three Stokes parameters for linearly polarized 
light  using the formal solution to the equation of transfer \citep{Miha78}.

\begin{equation}
\label{eq1}
\left\{
\begin{matrix}
L(\varepsilon)  \\
Q(\varepsilon)  \\
U(\varepsilon) 
\end{matrix}
\right\}
= \int dV \kappa(\varepsilon,{\bf r}) S(\varepsilon,{\bf r}) e^{-\tau(\varepsilon,{\bf r})}
\left\{
\begin{matrix}
1 + \cos^2\chi  \\
\sin^2\chi \cos(2\gamma) \\ 
\sin^2\chi \sin(2\gamma)  
\end{matrix}
\right\}
\end{equation}

\noindent where $\varepsilon$ is the photon energy, $S(\varepsilon,{\bf r})$,  is the source function, $\kappa(\varepsilon,{\bf r})$
is the opacity,  $\tau(\varepsilon,{\bf r})=\int{\kappa(\varepsilon,{\bf r}) d\zeta}$ is the optical
depth from a point ${\bf r}$ to a distant viewer,
$\chi$ is the scattering angle and $\gamma$ is the angle between the scattering 
plane and a reference direction on the sky.  Here and in what follows we describe the Stokes parameters in terms of the
luminosity seen by a distant observer, i.e. the total energy (in ergs s$^{-1}$ sr$^{-1}$) radiated by the
system in that direction.  Equation \ref{eq1} defines the scattered luminosity, $L$
(polarized plus unpolarized); observations are also affected by an
unscattered component, $L_u(\varepsilon)=L_0(\varepsilon)e^{-\tau(\varepsilon,{\bf r_x})}$
where ${\bf r_x}$ is the position of the X-ray source.  For the purpose of evaluating these
quantities we adopt a cylindrical coordinate system with the X-ray source at the origin 
and the viewing direction along the $\hat{\bf y}$ axis.  The angle $\gamma$ corresponds 
to the azimuthal angle on the plane of the sky and is measured relative
to a line which is the intersection between the orbital plane and the plane of the sky.
In the case of pure electron scattering the source function is $S=L_0/(4\pi r_x^2)$ where 
$r_x$ is the distance from the X-ray source and  the opacity is $\kappa(\varepsilon,{\bf r})=n({\bf r}) \sigma_{Th}$.

Some useful results are presented in the Appendix.  The fractional polarization relevant to observation is
$P=\sqrt{(Q^2+U^2)}/(L+L_u)$.  This quantity is zero when the system is viewed at inclination
$i=\pi/2$ and at either conjunction, i.e. orbital phase 0.5 or 1, $\Theta_V=0$ or $\pi$.
The polarization at $i=\pi/2$ is a maximum at quadrature, phase 0.25 or 0.75.
We can also define the polarization of the scattered radiation only,
i.e. $P_s=\sqrt{(Q^2+U^2)}/L$, and the value of this quantity at quadrature and $i=\pi/2$ depends only on
the orbital separation and on the distribution of gas density in the wind.
The value of $P$ is less than the value of $P_s$ by a factor proportional to the Thomson depth 
whenever the X-ray source is not eclipsed. 
At $i=0$, i.e. when the system is viewed face-on, the polarization rotates with orbital phase at a rate
twice the orbital rotation rate.

As an illustration of these results, we calculate the polarization produced by a  single-scattering
calculation by numerically evaluating equation \ref{eq1}.  The wind velocity is assumed to be radial
relative to the star with a speed given by $v(r)=v_0+(v_{\infty}-v_0)(1-r_{*}/r)$
where the terminal velocity is $v_{\infty}$=1000 km s$^{-1}$ and $v_0$=100 km s$^{-1}$, and
$r_{*}=10^{12}$ cm.  The wind mass loss rate is $\dot M_{wind}=10^{-8} M_{\odot} {\rm yr}^{-1}$.
The orbital separation is $a=1.5 r_{\*}$.

In the remainder of this paper we discuss quantitative calculations of polarization produced by wind scattering.  These
include calculations based on both the very simple spherical wind with parameters given in the previous paragraph
and also on hydrodynamic calculations of the wind density structure which make no assumptions about spherical symmetry.
In the former case, it is illuminating to estimate the accretion rate onto the compact object if the accretion is supplied
solely by the wind.  This can be done using simple estimates based on  \citet{Bond44}  and \citet{Davi73}, i.e.
$\dot M_{acc}=\pi R_{acc}^2 n_x v_x m_H$ where the accretion radius is $R_x=2GM/v_x$, $M$ is the mass of the compact
object and $v_x$ is the speed of the wind at the X-ray source.  This corresponds to
$\dot M_{acc}/\dot M_{wind}= 1.8 \times 10^{-4} (M/M_\odot)^2 v_{x8}^{-4} a_{12}^{-2}$  where $v_{x8}=v_x/(1000 {\rm km s}^{-1})$,
 $a_{12}=a/(10^{12} {\rm cm})$ or an X-ray luminosity $L_x\simeq 1.8 \times 10^{35} {\rm erg s}^{-1}$ using the
parameters given in the previous paragraph and assuming an efficiency of converting accreted mass into energy of 0.1.
This luminosity is less than the time averaged luminosities of most HMXBs by factors $\sim$5 -- 100, which likely reflects
the fact that wind mass loss rates may be greater, additional mass can be supplied by Roche lobe overflow, and the
wind and accretion flow dynamics can be enhanced by the influence of the compact object gravity and ionization.
The results in what follows, those which are based on the simple spherical wind approximation, must
interpreted subject to this caveat.

An additional caveat which applies to essentially all of the results presented in this paper is the assumption of single
scattering.  The validity of this assumption depends on the wind optical depth from the X-ray source;
multiple scattering is important when this quantity approaches or exceeds unity.  For parameters similar
to those discussed so far, this quantity for a distant observer corresponds to
$\tau_{Th} \sim n_x a \simeq 2 \times 10^{-4} \dot M_8 a_{12}^{-1} v_{x8}^{-1}$,
where $\dot M_8= \dot M/(10^{-8}M_{\odot}{\rm yr}^{-1})$.  Comparing this expression
with the accretion luminosity estimate given above implies that multiple scattering can be important when
the accretion luminosity exceeds $\sim10^{38} {\rm erg s}^{-1}$ for a 1 $M_\odot$ compact object.   The likely effect of
multiple scattering is  to produce smaller net polarization than for single scattering, since the second and subsequent
scatterings have a wider range of scattering angles and planes than for single scattering.  If the
depth is moderate, i.e. $\leq$ 10, there will still be a significant fraction of photons which reach the
observer after a single scattering.  If so the net polarization will likely be less than predicted here,
by factors of order unity, and our results should be modified to include multiple scattering effects for such sources.  

\begin{figure*}[p] 
\includegraphics*[angle=0, scale=0.5]{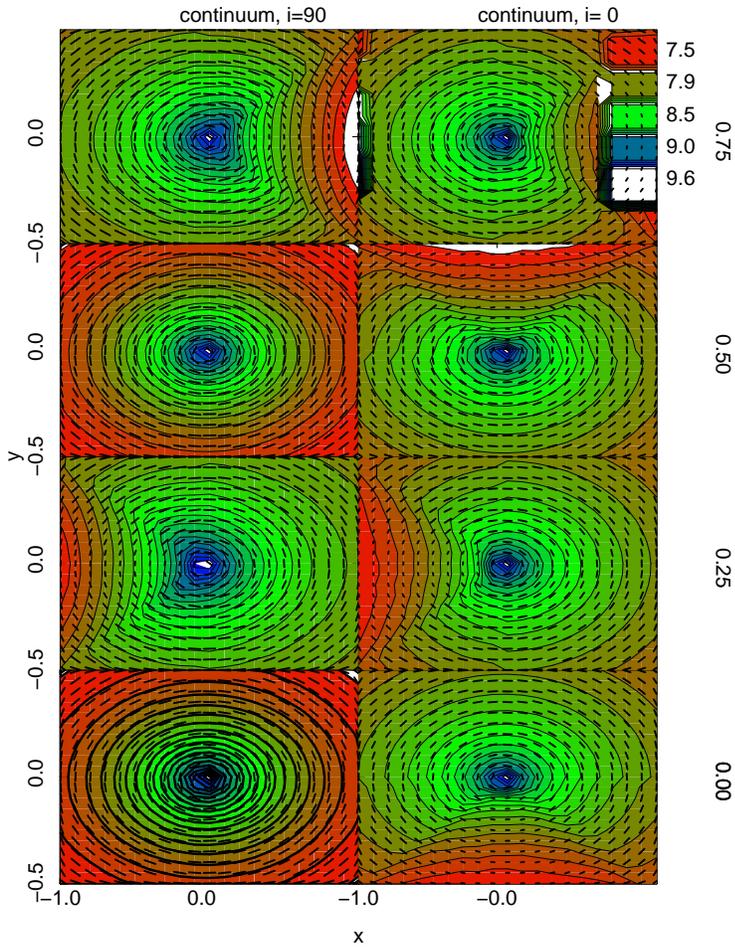}
\caption{\label{fig2} Map of intensity and polarization vectors projected on the sky
  for four orbital phases and two inclinations.  Colors correspond to scattered intensity
  according to the color bar, in which the labeled quantities correspond to log$_{10}$ of the specific
  intensity.  Short black lines correspond to polarization vectors.
  Spherically symmetric wind is assumed.} 
\end{figure*}

Maps of polarization projected against the sky
are shown in figure \ref{fig2}.  This shows contours of constant intensity
as solid colors separated by solid black curves, along with lines corresponding to polarization vectors,
which appear as dashed curves.  These are plotted vs position in units of $10^{12}$ cm.
These illustrate many of the results presented in the Appendix:  At high inclination, $i=\pi/2$,
and at orbital phase angles 0 and $\pi$ (orbital phases 0 and 0.5) polarization vectors are perfectly
circumferential, contours of constant intensity are also circular and the net polarization is zero.
At orbital phase angles $\pi/2$ and $3\pi/2$ (orbital phases 0.25 and 0.75) the net polarization is
maximum.  
At inclination $i=0$ the star always influences the polarization,
so that the net polarization fraction is constant, but the position rotates with the orbit.

\begin{figure*}[p] 
\includegraphics*[angle=90, scale=0.3]{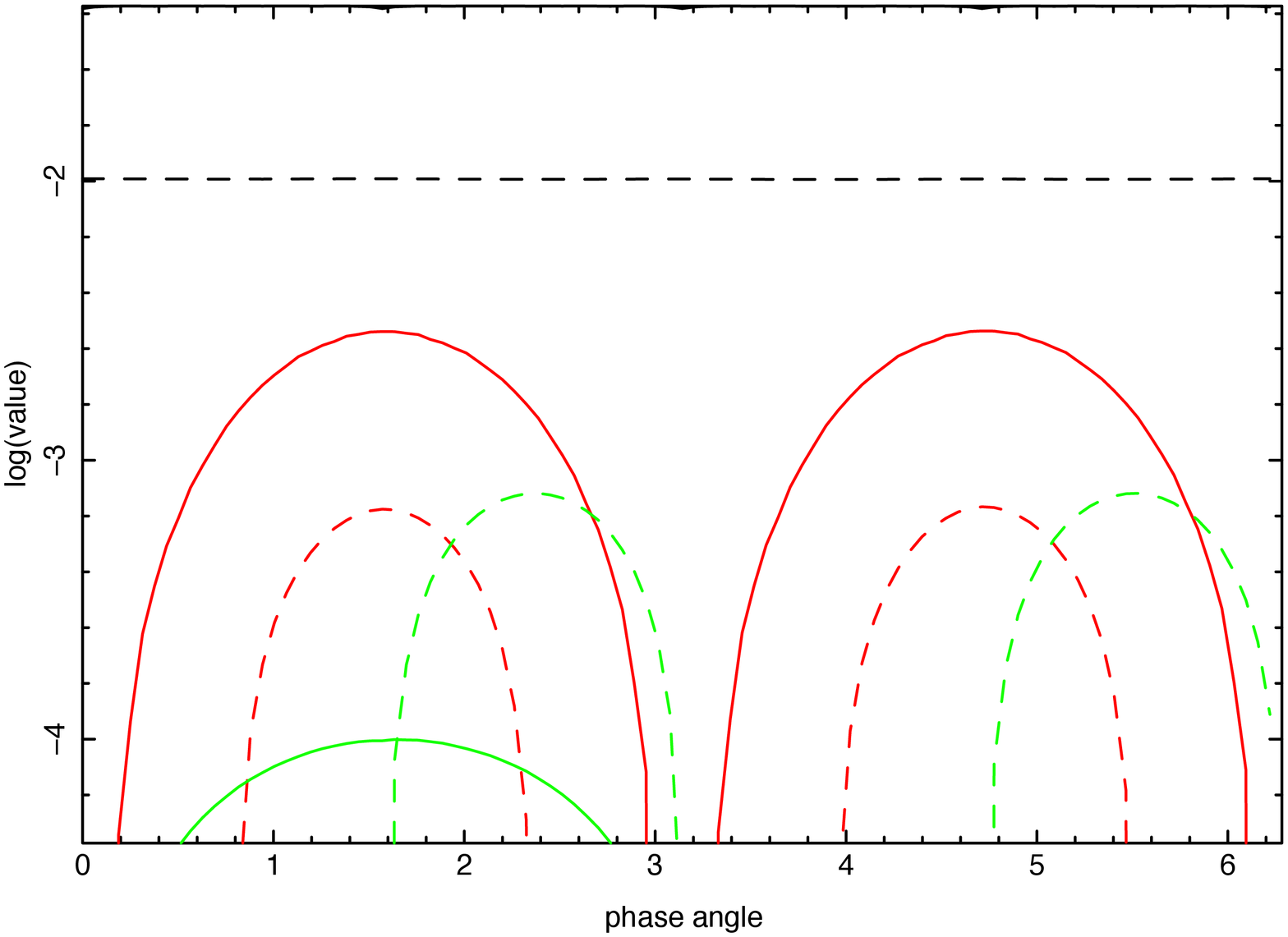}
\caption{\label{fig3} Stokes parameters for spherical wind, Q (red ) and U (green)  vs. phase
  for inclinations $i=0$ (dashed)  and $i=\pi/2$ (solid).  The total intensity I  is shown as dashed black for
  $i=0$.  At $i=0$ the polarization oscillates between
Q and U along orbital phase, and the vector sum is constant.
At $i=\pi/2$ U is always small (in this convention) and Q is a maximum 
at phase angles $\pi/2$ and $3\pi/2$.}
\end{figure*}

\begin{figure*}[p] 
\includegraphics*[angle=90, scale=0.4]{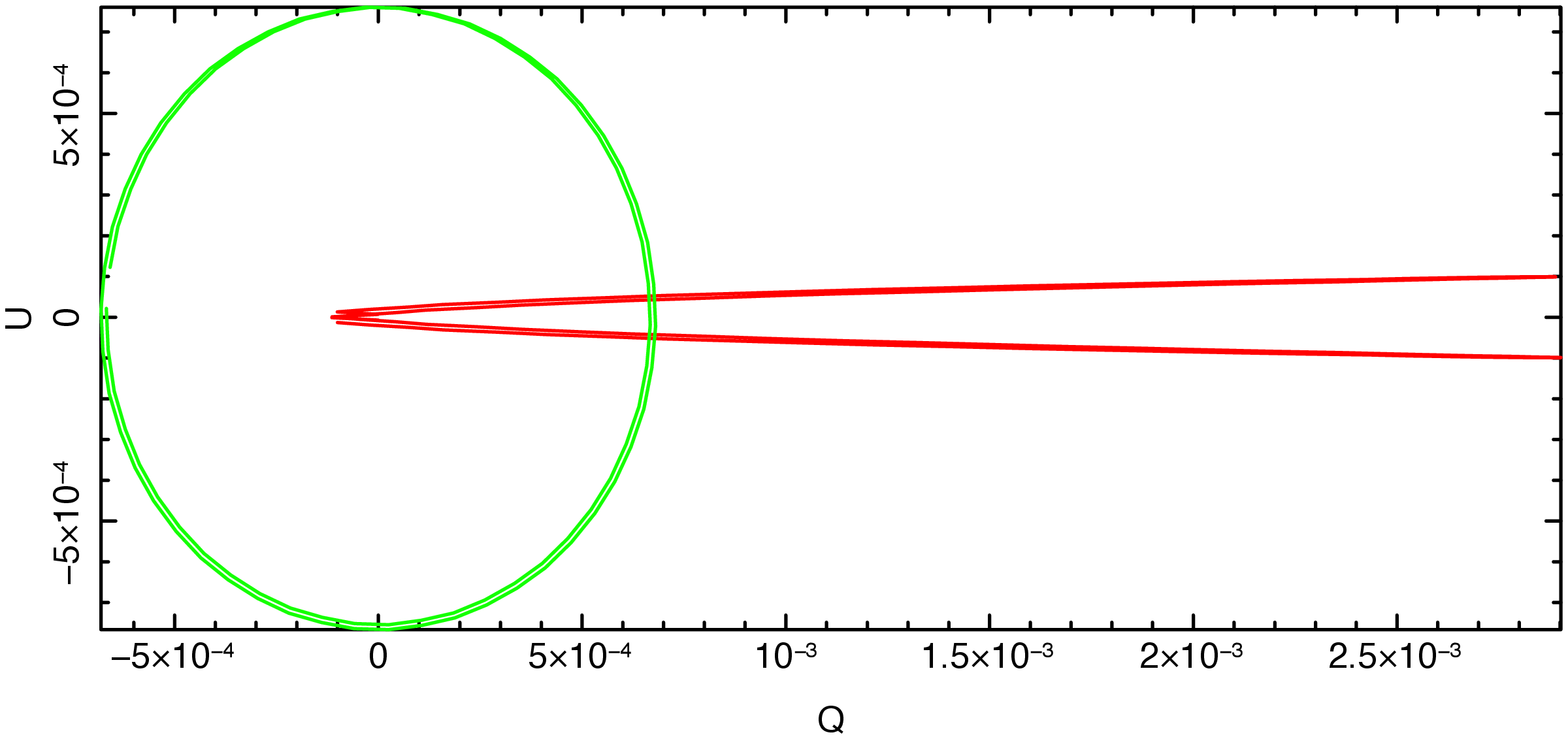}
\caption{\label{fig4} Stokes parameters for spherical wind, Q vs U $i=0$ (green)
  and $i=\pi/2$ (red).  This again illustrates the circular path of the net polarization
direction with orbital phase at $i=0$ and the constancy of the direction at $i=\pi/2$.}
\end{figure*}

More insight comes from the Stokes quantities as a function of the orbital phase.  These
are shown in figure \ref{fig3} for inclinations $i=0$ and $i=\pi/2$.  U is green while Q is shown
in red; solid is $i=\pi/2$ and dashed is $i=0$.  This further illustrates the results from the
previous figure:  at $i=0$ the polarization oscillates between Q and U along orbital phase, and the
vector sum is constant.  At $i=\pi/2$ U is always small (in this convention) and Q is a maximum
at phase angles $\pi/2$ and $3\pi/2$.
Another way of displaying the same thing is shown in
figure \ref{fig4}, which shows U plotted vs Q, for $i=0$ in green, and $i=\pi/2$ in red.  This
shows that the trajectory of U vs. Q is circular at $i=0$ and becomes linear at $i=\pi/2$.  As shown by
\citet{Brow78}, this trajectory is an ellipse for intermediate inclinations, and the cases
in figure \ref{fig3} are the extremes of eccentricity.   This has been suggested as a means
for measuring inclination \citep{Rudy78}.   

\begin{figure*}[p] 
\includegraphics*[angle=270, scale=0.4]{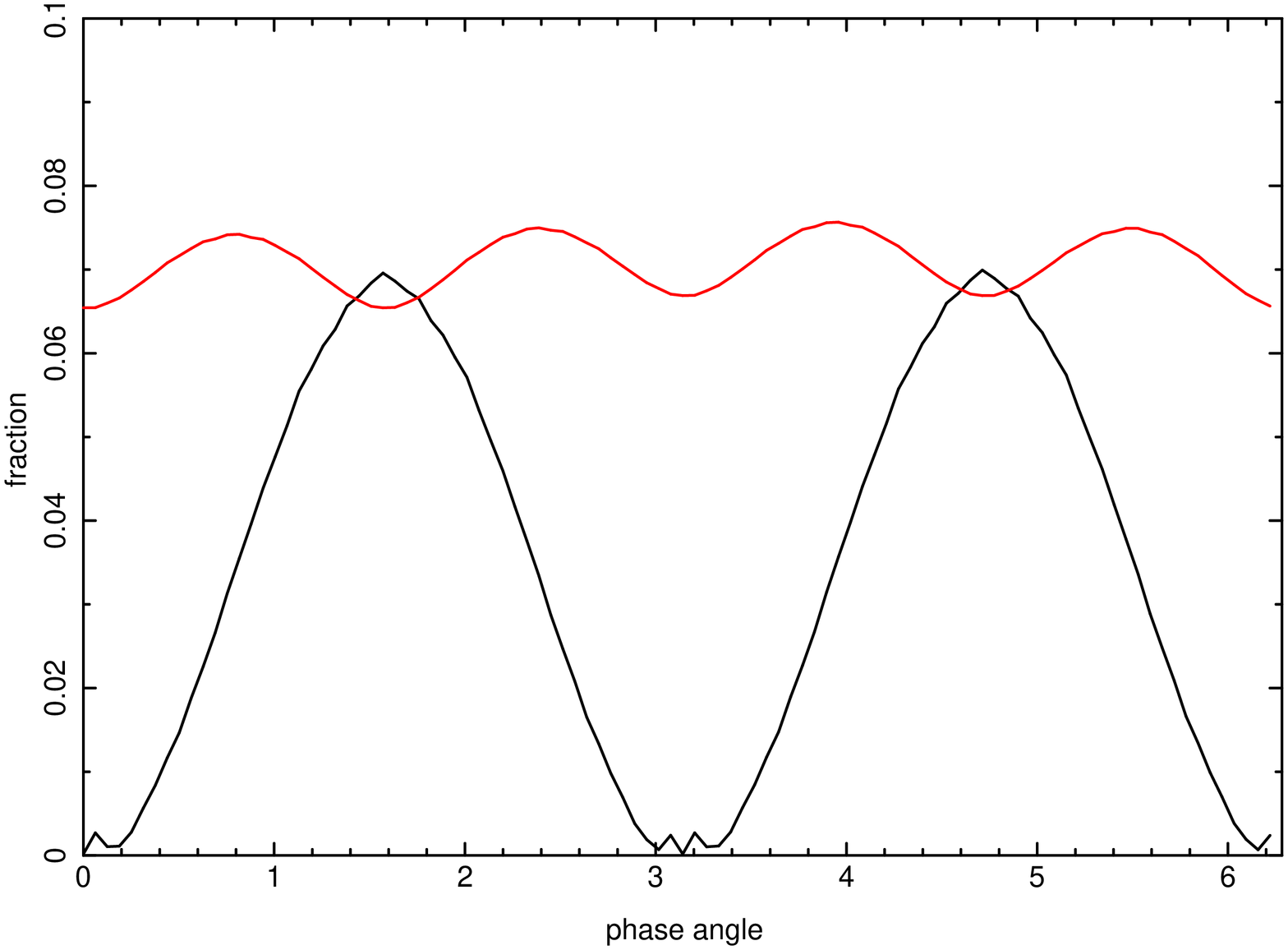}
\caption{\label{fig5}  Polarization fraction of scattered radiation i.e.
  $P_s=\sqrt{(Q^2+U^2)}/L$ for spherical wind, $i=0$ (red)
  and $i=\pi/2$ (black).}
\end{figure*}

\begin{figure*}[p] 
\includegraphics*[angle=270, scale=0.4]{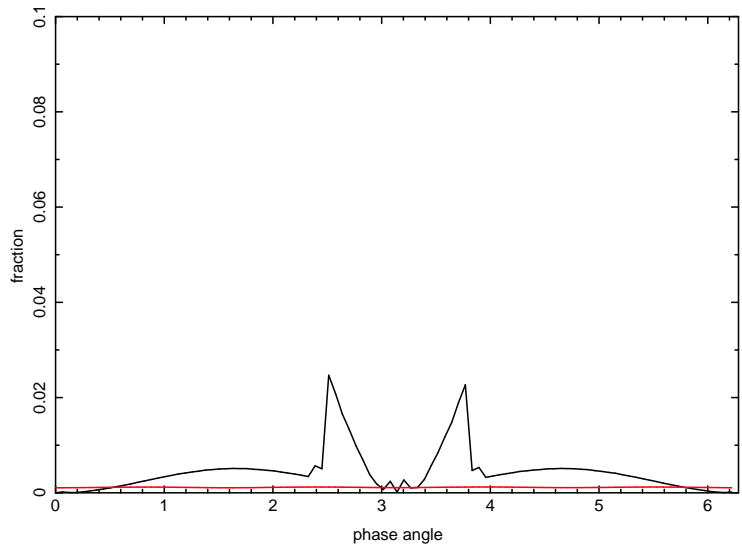}
\caption{\label{fig6}  Polarization fraction relative to total radiation, scattered plus direct,
  for spherical wind, $i=0$ (red)
  and $i=\pi/2$ (black).}
\end{figure*}

\begin{figure*}[p] 
\includegraphics*[angle=270, scale=0.5]{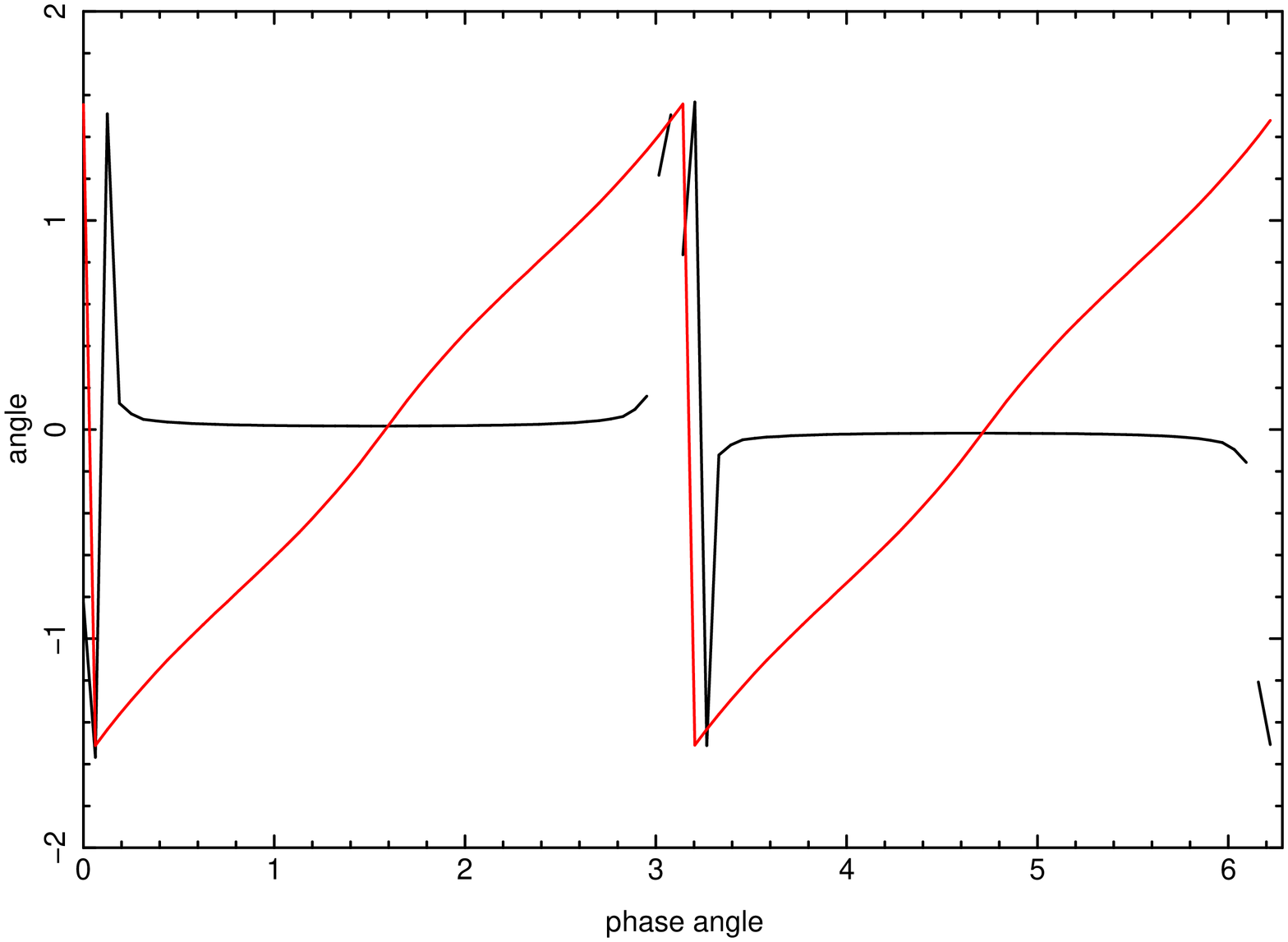}
\caption{\label{fig7} Polarization angle for spherical wind, $i=0$ (red)
  and $i=\pi/2$ (black).}
\end{figure*}

Figure \ref{fig5} shows the scattered polarization fraction vs. orbital phase,
i.e. $P_s=\sqrt{(Q^2+U^2)}/L$.  This shows that the
maximum polarization fraction is $\simeq$10$\%$ for the parameters chosen here, when the
polarized component is compared with the total scattered component.  This is comparable to the
result in the Appendix, calculated for a thin shell at the star rather than for an extended wind.
We emphasize that this value depends primarily on geometric quantities:  the extent of the wind, and
orbital separation relative to the stellar radius.  It does not depend on the wind
density or column density since it is a comparison of scattered quantities.  The difference between
$i=0$ and $i=\pi/2$ is clearly apparent:  the former  produces approximately constant polarization fraction
and the latter oscillates between 0 and a value comparable to the $i=\pi/2$ value.
Figure \ref{fig6} shows the polarization fraction measured relative to the total radiation, scattered
plus direct, i.e. i.e. $P=\sqrt{(Q^2+U^2)}/(L+ L_u)$.  This illustrates the diluting effect of the direct
radiation for this model, which has $\tau_{Th}\simeq 0.25$.  Maximum linear polarization at $i=\pi/2$occurs
just following the eclipse transition, where the departures from circular symmetry on the sky are not negligible,
and where the direct radiation is blocked by the primary star.  At mid-eclipse the polarization at $i=\pi/2$ is
zero due to symmetry.
Figure \ref{fig7} shows the polarization angles from the same set of models.  The
angle sweeps through 180 degrees twice per orbital period for $i=0$, while the
angle is constant (though undefined near conjunctions) for $i=\pi/2$.

\section{Resonance Line Scattering: Single Line}
\label{sec3}

The models discussed in section \ref{sec1} include solely electron scattering;
they do not include resonant scattering in bound-bound transitions.  This
process (when associated with UV and optical transitions) is the dominant driving
mechanism for the winds from early type stars.  Here we consider
scattering in the X-ray band.   In the absence of X-ray ionization, the ions
which are most abundant in early-type star winds are in charge states from
$\sim$1 -- 5 times ionized, and do not have many strong X-ray resonance
line transitions.   Ionization by a compact X-ray source can produce
ions of arbitrary charge state, depending on the X-ray flux and gas density,
and so resonance scattering can affect the scattered  X-ray intensity and
polarization.  The ionization structure in HMXB winds has been explored by
\citet{Hatc77} who showed that the surfaces of constant ionization in a spherical
wind are either nested spheres surrounding the X-ray source, or else open surfaces
enclosing the primary star.  The ionization distribution depends on a parameter
$q=\xi/\xi_x$ where $\xi_x=L_x/(n_xa^2)$ where $L_x$ is the ionizing X-ray luminosity,
$n_x$ is the gas density at the X-ray source and $a$ is the orbital separation.
This is a particular example
of the scaling of ionization in optically thin photoionized gas, which depends on
the ionization parameter $\xi=L_x/(n(r) r_x^2)$ \citep{Tart69,KallmanBautista01}.
The properties of resonance scattering as applied to X-ray polarization
in HMXBs are dominated by the fact that,
owing to the restricted regions where ionization is suitable,
most lines can only have strong opacity
over a fraction of the wind.  The regions where this occurs most often resemble
spherical shells surrounding the X-ray source.

Resonance scattering differs from electron scattering in its sensitivity to the velocity
structure of the wind.  At a given observed photon energy, scattering can occur only
over a resonant surface with shape determined by the wind velocity law; in a
spherical wind with monotonic velocity law these surfaces are open surfaces of revolution
symmetric about the line of sight to the X-ray source.

Resonance scattering affects polarization by 
redistributing the polarization among the components
of the Stokes vector according to the phase matrix which is a linear
 combination of the Rayleigh and isotropic  phase matrices; coefficients 
of the matrix depend on the angular momentum quantum 
numbers of the initial and final states of the transition, $j$ \citep{LeeBlandfordWestern94}.  
The matrices describing this process have been calculated by \citet{Hamilton47}.  
In practice most of the strongest X-ray resonance lines have initial and final values
$j_{lower}=1/2$ and $j_{upper}=3/2$, which correspond to the Rayleigh phase matrix,
and we adopt this in our calculations.  Similar assumptions were employed in 
\citet{DoraKall10}.

Polarization also depends on the scattering angle, and the resonance
condition constrains the scattering geometry, thereby imposing a relation
between the energy of the scattered photon  and its polarization.
Resonance scattering can have a cross section which is much  greater than
electron scattering, but only over the relatively narrow energy band spanned by the
resonance line, including the effects of Doppler broadening associated with the
wind motion.  Within this band, the polarization can be greater than that produced by
electron scattering, owing to the greater cross section, given suitable geometry.  

In the remainder of this section we illustrate these effects by generalizing
the single scattering spherical wind calculations to include resonance
scattering.  We consider a single resonance line, chosen to crudely resemble
a line such as O VIII L$\alpha$.  The parent ion is assumed to exist over
a range of ionization parameter $1\leq {\rm log}(\xi) \leq 3$.
We calculate the optical depth using the Sobolev expression \citep{Castor}:

\begin{equation}
\label{equation2}
  \tau_{line}(\varepsilon,{\bf r})=\frac{\pi e^2}{mc} f n({\bf r})x_{ij}y_j \frac{\lambda r}{v(r)(1+\frac{z^2}{r^2}\left(\frac{d{\rm ln}v(r)}{d{\rm ln}r}-1\right)}
  \end{equation}

\noindent where $f$ is the oscillator strength, $n({\bf r})$ is the gas number density, $x_{ij}$ is the ion fraction, $y_j$ is the
element abundance, $z$ is the position along the line of sight, and $\lambda$ is the line wavelength.  For the source
function we adopt the same expression used for electron scattering $S=L_0/(4\pi r_x^2)$ since thermalization is unimportant and the
size of the X-ray source is small compared with the other length scales of interest for HMXBs.

%
%
%

\begin{figure*}[p] 
\includegraphics*[angle=270, scale=0.5]{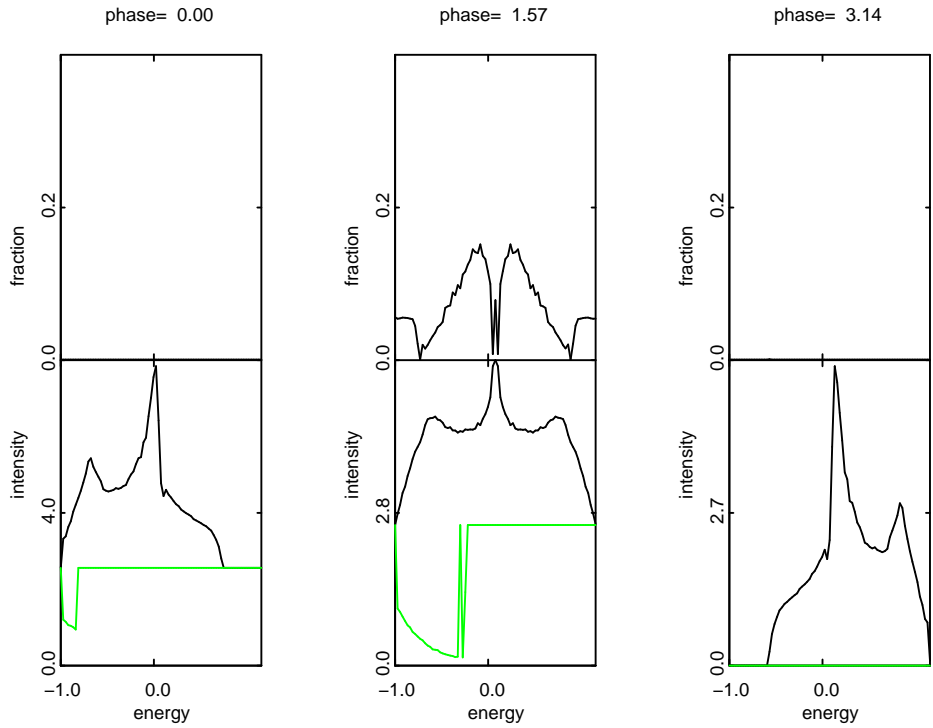}
\caption{\label{fig8} Spectrum, polarization fraction and angle in a resonance line vs. energy
  in units of the terminal velocity at orbital phase 0.25 for inclination $i=\pi/2$}
  \end{figure*}

Results of our simple single line calculation are shown in figure \ref{fig8}.
This shows the line profile as a function of orbital phase for a system viewed edge on ($i=\pi/2$).
For each phase we display the luminosity, with the transmitted luminosity shown in black and the
scattered luminosity shown in green (bottom panel) and polarization fraction (top panel).
The polarization angle is not shown because there is no significant dependence on energy
or orbital phase when the system is viewed at high inclination; the polarization is always
perpendicular to the orbital plane.  The horizontal axis is energy in units of the wind terminal
velocity.

At phase 0 (superior conjunction of the X-ray source) the shape of the profile in
luminosity is similar to P-Cygni profiles familiar from UV resonance
lines in hot stars:  the outflow produces blue-shifted absorption (shown as negative energy in these
figures) of the continuum.  This absorption is offset from the zero of energy owing to the fact that
the continuum source in this case is the compact X-ray source, and the wind speed at the X-ray source
is approximately half the terminal speed.  In addition, the X-ray source creates an ionized region
within which line scattering cannot occur.  As a result,
the absorption occurs in a relatively narrow region of energy near the energy corresponding to the
wind terminal velocity.  The scattered emission, on the other hand, is essentially symmetric in energy, since
scattered emission comes from a region which is not necessarily along the line of sight to the X-ray source.  
Projection effects make the scattered emission appear at all energies $|\varepsilon| \leq \varepsilon_0 v_\infty/c$
where $\varepsilon_0$ is the line rest energy.  The scattered emission is unpolarized at phase 0 for the same
reason as in the pure electron scattering case.
At phase 0.25 (quadrature) the wind is viewed perpendicular to its velocity vector at the X-ray source,
and therefore the absorption covers a large fraction of the energy spanned by the wind.
The scattered emission is polarized up to $\sim$50$\%$; the degree of polarization is symmetric around
the center of the line, owing to the fact that the wind velocity structure is symmetric around the line
of centers.  At phase 0.5 (inferior conjunction of the X-ray source)
there is no transmitted flux (due to occultation by the star) and the emission is predominantly red-shifted
owing to the location of the X-ray ionized zone in the receding part of the wind.  In this case,
even though the flux is all scattered, the polarization fraction is negligible owing to the circular
symmetry of the scattering region of the wind as viewed in the plane of the sky.

\section{Absorption Effects}

The ionization balance in an HMXB wind is determined at each point by the
X-ray flux and by the gas density.   The X-ray flux in turn depends
on the effects of geometrical dilution and on attenuation. 
In order to do so, we utilize results from {\sc xstar} \citep{KallmanBautista01} at each point
in the wind in order to determine the opacity.   This is done along
radial rays originating from the X-ray source, so that the opacity
at each point is calculated using an ionization parameter which corresponds
to the flux transmitted along the corresponding ray.

This single stream
transfer treatment is similar to the transfer treatment used by  {\sc xstar}
with one difference:  {\sc xstar} uses the local spectral energy distribution (SED)
to calculation the ionization along a ray; here we use the same unattenuated
SED everywhere for the ionization calculation (so we can use a stored table),
but we calculate the ionization parameter self-consistently, i.e. by calculating the 
transmitted flux at each position and then using that value to calculate the ionization parameter.   
This simplification allows for more efficient computation without unduly sacrificing physical realism.
This approximation will be least accurate when the column density is highest, i.e.
$\geq 10^{23}$ cm$^{-2}$.  In such regions, the ionization is expected to be low,
owing to attenuation, and X-ray scattering will be relatively unimportant.  Our
approximation will tend to produce less attenuation at a given column density
than a self-consistent calculation would.

\section{Resonance Line Scattering:  Ensemble of Lines}
\label{sec3b}

\begin{figure*}[p] 
\includegraphics*[angle=90, scale=0.5]{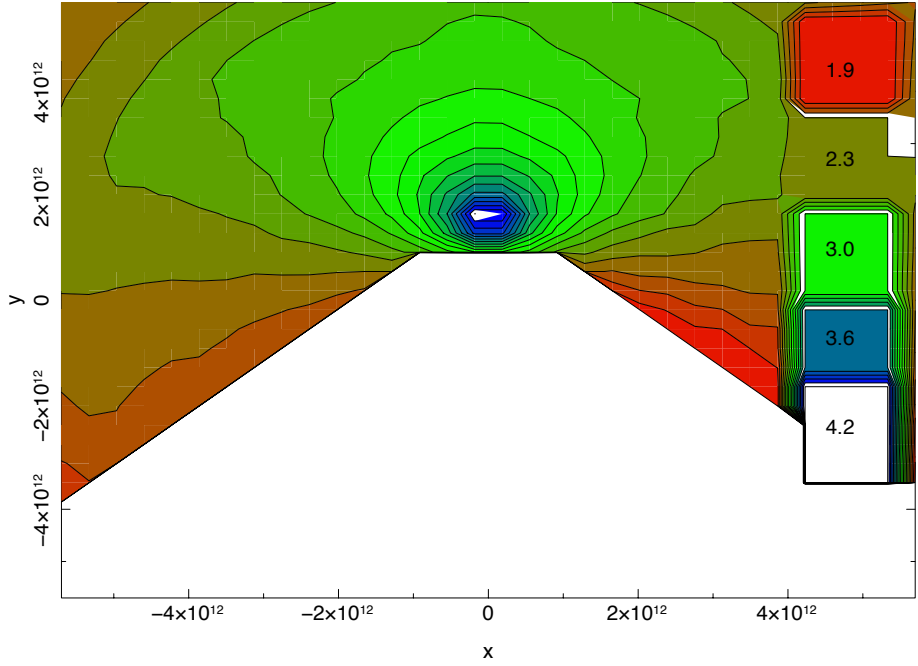}
\caption{\label{fig9}Distribution of ionization parameter in the orbital plane
  for a spherical wind with $\dot M_{wind}=10^{-5} M_\odot {\rm yr}^{-1}$
   illuminated by a 10$^{38}$ erg s$^{-1}$ X-ray source.  Contours are
  labeled with log$\xi$.}
\end{figure*}

\begin{figure*}[p] 
\includegraphics*[angle=270, scale=0.5]{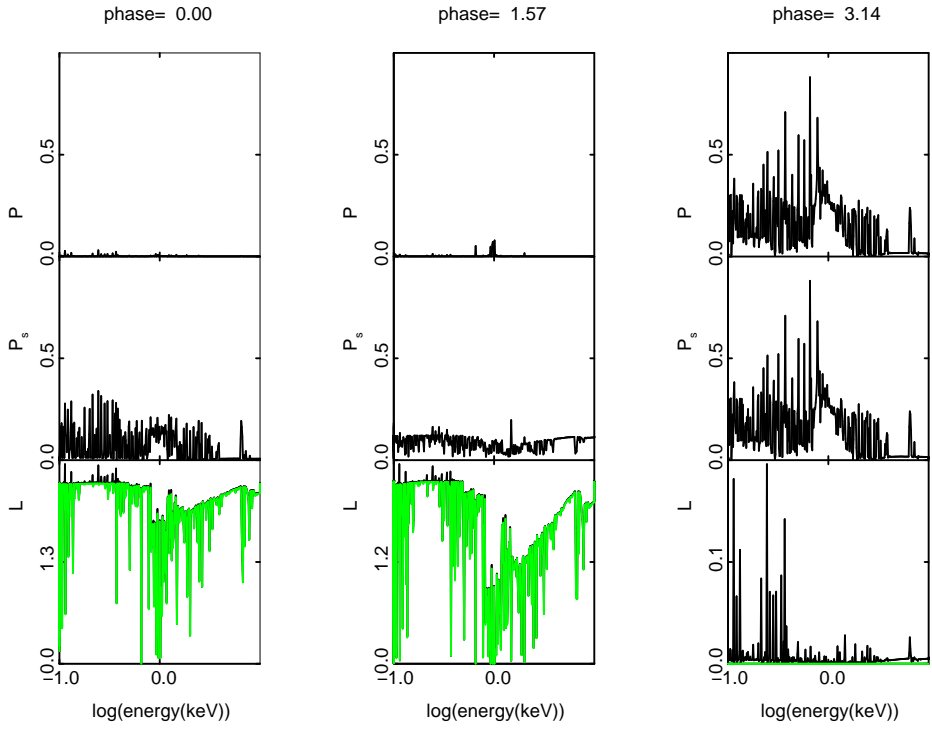}
\caption{\label{fig10} Spectrum and polarization fraction vs energy
for spherical wind in units log(E/keV) at orbital phase 0.25 for inclination $i=\pi/2$}
  \end{figure*}

X-ray ionization creates an ensemble of resonance lines in an HMXB wind from the many
trace elements such as C, N, O, Ne, Mg, Si, S, and Fe.  At
any point in the wind the ionization depends most sensitively on the ionization parameter,
defined in the previous section.  Figure \ref{fig9} shows the distribution of ionization
parameter produced in a wind with a density distribution which is spherically symmetric
around the primary star, as discussed so far. 
We adopt a spherical wind with mass loss rate $10^{-5} M_\odot {\rm yr}^{-1}$, corresponding to a
total wind column density of $3 \times 10^{23}$ cm$^{-2}$, and an X-ray source luminosity of $10^{38}$ erg s$^{-1}$.
Each spatial region with a distinct ionization parameter will have a different distribution of ion abundances,
and hence a unique distribution of resonance line opacity.  These can be calculated
under the assumption of ionization equilibrium.  Figure \ref{fig10} 
show examples of the spectrum and the polarization fraction produced by
such a distribution from the spherical wind shown in figure \ref{fig9}.
Ionization and line opacities were calculated as a function
of ionization parameter using the {\sc xstar} code \citep{KallmanBautista01}.
These were binned into 1000 energy bins over energy from 0.1 eV to 10$^4$ eV.
Since the Doppler shifts associated with the wind speed is less than the energy bin size,
we adopt an approximate form for the source function each resonance line:

\begin{equation}
  S(\varepsilon,{\bf r})=\frac{L_0}{4\pi r_x^2}  \frac{v_{\infty}-v_0}{\Delta \varepsilon c/\varepsilon}
\end{equation}

\noindent where the quantity
$(v_{\infty}-v_0)/(\Delta \varepsilon c/\varepsilon)$ takes into account the fact that the line
does not conver the entire energy bin.  This treatment assumes that the line optical depths are not large,
which is an adequate approximation for our situation.  
Depolarization effects at large Sobolev optical depths associated with multiple scatterings are not taken into account.
These  calculations serve to illustrate the magnitude of the polarization effects expected from
resonance scattering in HMXBs.  Simulations suitable for quantitatively diagnosing the wind or the X-ray source properties
will require reexamination of these assumptions.

Results of spectra and polarization fractions are shown in figure \ref{fig10}.
These demonstrate that the resonance lines cover a significant fraction of the X-ray energy
band and that they can scatter a significant flux of X-rays, and create polarization fractions
as high as 0.5, much greater than would be produced by electron scattering alone.
A difference between resonance scattering and electron scattering is that resonance
scattering in a given line occurs within a relatively narrow spatial region where the parent
ion is most abundant.  For X-ray lines, these are most likely to be approximately spherical
surface surrounding the X-ray source.  This region tends to produce a small polarization which
is almost independent of the viewing angle.   Thus resonance scattering produces
weaker modulation  of the polarization with orbital phase than does electron scattering.

It is also worth noting that the effects of scattering can be either polarizing or not polarizing,
depending on the scattering angle.  Also, lines which appear in absorption in the
spectrum can have enhanced polarization in their troughs owing to the presence of scattered light
in the residual intensity.  In the spectrum shown in figure \ref{fig10} the polarization
in the absorption lines is generally lower than in the emission features, or in the adjacent
continuum.  This is due to the fact that the scattered light in the residual intensity is
forward scattered, and so is not polarized because of the scattering angle.  

\section{Hydrodynamic Models}
\label{sec4}

The stellar wind in HMXBs is not spherically symmetric. Physical processes 
which affect the wind and accretion flow in HMXBs include: the three-dimensional 
geometry of the flow (i.e. the flow both in and out of the 
orbital plane), rotational forces, the influence of gravity and radiation 
pressure from both stars in the binary, transport of the radiation 
from the stars into the flow and the reprocessed radiation out of the 
flow, X-ray heating and ionization, and departures from thermal equilibrium 
due to advection and adiabatic heating and cooling.  The dynamics
of X-ray heated winds have been discussed by \citet{Fran80}, and \citet{Hatc77}, 
and multi-dimensional models have been 
calculated by \citet{Blon90, Blondin94, Mauc08}

\begin{figure*}[p] 
\includegraphics*[angle=0, scale=0.4]{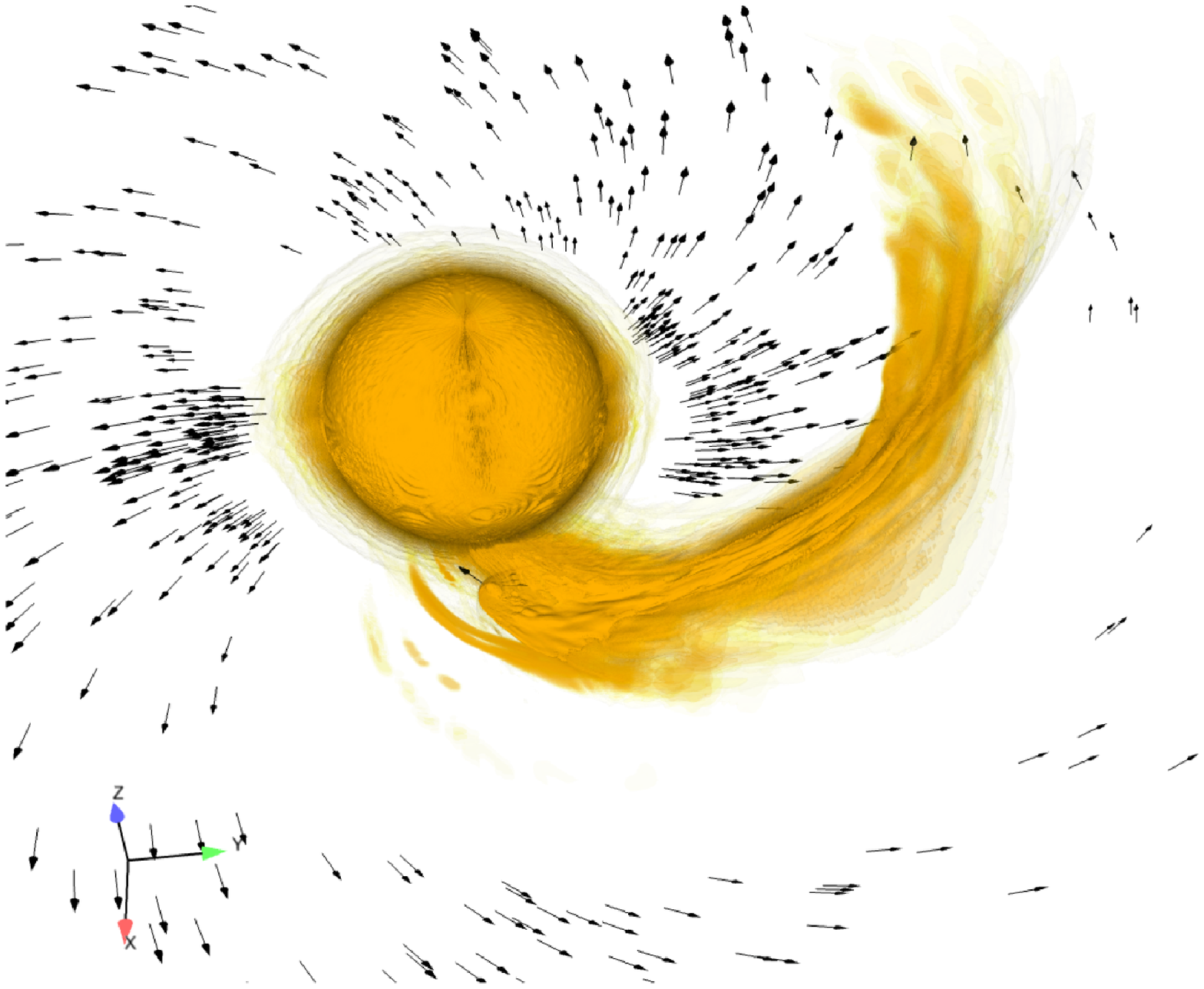}
\includegraphics*[angle=0, scale=0.4]{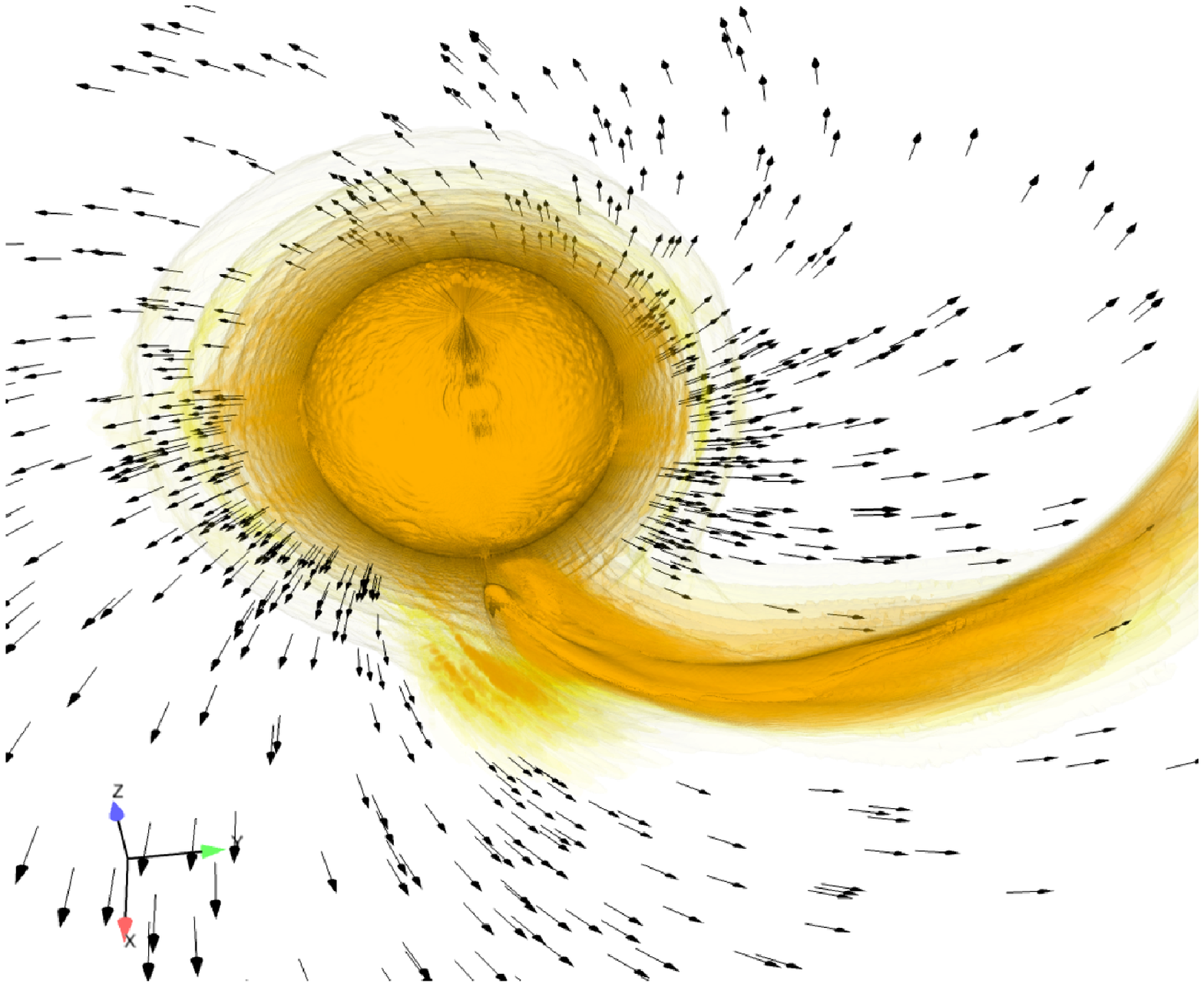}
\caption{\label{fig11} Density contour and vector plot of wind hydrodynamic
  models.   Right panel is model with
  $\dot M=1.7 \times 10^{-6} M_\odot {\rm yr}^{-1}$
  left panel is model with
  $\dot M=4.0 \times 10^{-7} M_\odot {\rm yr}^{-1}$.  Spatial scale is
  the same for both panels.  Yellow contour corresponds to a density of
  $6 \times 10^8 {\rm cm}^{-3}$. }
\end{figure*}

In this section we illustrate the effects of the wind hydrodynamics on the 
polarization. We use two sample wind hydrodynamic models calculated using 
a numerical model similar to that described in \citet{Blon09}
but extended to three dimensions.  The hydrodynamic models 
are computed on one hemisphere of a spherical grid, assuming reflection 
symmetry about the orbital plane.  A non-uniform grid of 448 (r) by 
128 ($\theta$) by 512($\phi$) zones is used with highest resolution near the 
surface of the primary star and in the vicinity of the accreting neutron 
star.  This model calculates the wind dynamics in three dimensions taking 
into account the radiative driving by the UV/optical radiation from 
the primary, and also the gravity of the star and the compact X-ray source. 
A fixed X-ray luminosity of 10$^{36}$ erg/s is used to calculate X-ray heating 
using the approximate formulae given in Blondin (1994) and the effects 
of X-ray ionization on the dynamics via changes in the UV radiation 
force multiplier.  The two models have identical dimensions:
primary radius 2.4 $\times 10^{12}$ cm and orbital separation  3.6 $\times 10^{12}$ cm.
They differ in their mass loss rates, which are 4 $\times 10^{-7} M_\odot {\rm yr}^{-1}$ and
1.7 $\times 10^{-6} M_\odot {\rm yr}^{-1}$.  The maximum wind speed in
both models is approximately 1600 km s$^{-1}$.
A plot showing the density contours and velocity vectors
is shown in figure \ref{fig11}.  This
clearly shows the influence of the gravity of the compact object in creating a region of higher
density and non-radial wind flow in the vicinity of the X-ray source.

The distorted stellar wind of the hydrodynamic models is illustrated with velocity 
vectors in the orbital plane and a series of transparent density isosurfaces.
Right panel is the model with a higher mass loss rate (Mdot = 1.7e-6); left panel
is a lower mass loss rate (Mdot = 4.0e-7).  The lowest density isosurface corresponds
to a density of 4e9 cm$^{-3}$ in both panels.  Velocity vectors are not shown for values
less than 800 km/s.

The high mass loss rate model has a denser spherical component of the wind, but a
less pronounced accretion wake.  The low mass loss rate model has a larger volume
of wind moving at low velocity due to photoionization of the wind in the vicinity 
of the X-ray source.  The result is
a denser wind coming off the primary along the line of centers of the binary system
and a larger, denser accretion wake that wraps more tightly around the primary star.

\begin{figure*}[p] 
\includegraphics*[angle=90, scale=0.5]{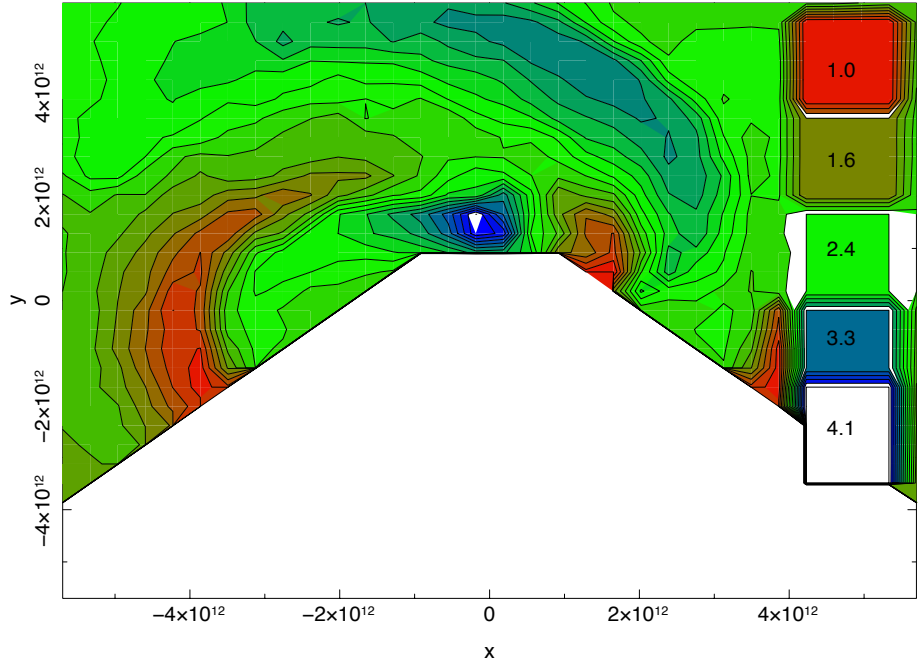}
\caption{\label{fig12}Distribution of ionization parameter in the orbital plane
  for hydrodynamic wind illuminated by a 10$^{36}$ erg s$^{-1}$ X-ray source.  Contours are  labeled with log$\xi$.}
\end{figure*}

We apply the same calculation of X-ray scattering to the hydrodynamic wind
as was done in section \ref{sec3}. We assume an X-ray source
luminosity of $10^{36}$ erg s$^{-1}$ and a $\Gamma$=2 power law ionizing spectrum.
The distribution of ionization parameter in the orbital plane is shown in figure
\ref{fig12}.  This shows a spiral structure, owing to a corresponding structure
in the gas density.

\begin{figure*}[p] 
\includegraphics*[angle=270, scale=0.5]{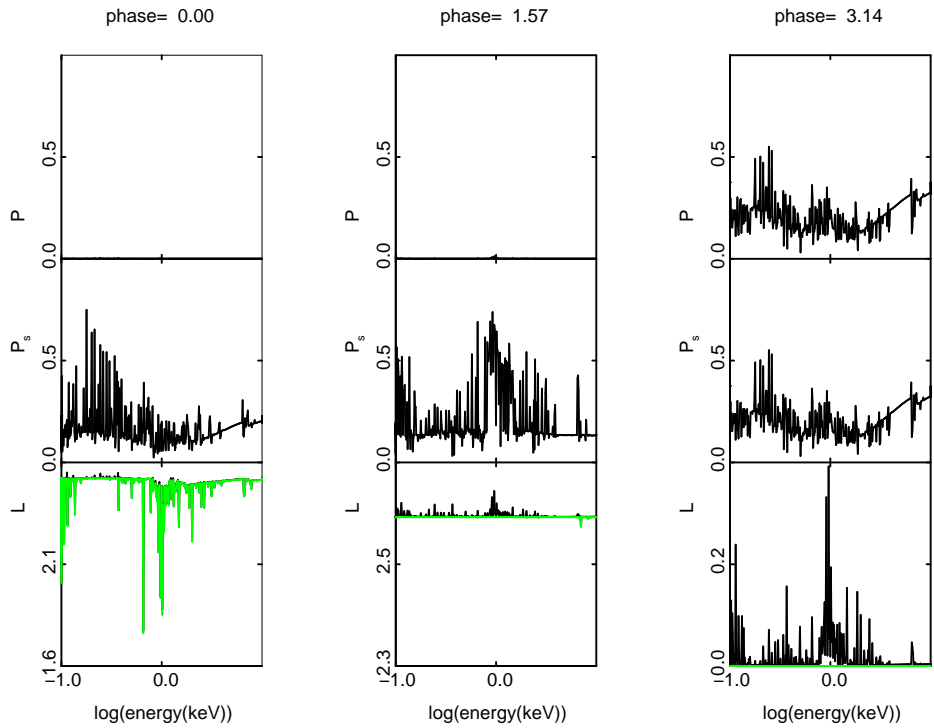}
\caption{\label{fig13}  Spectrum and polarization fraction vs energy
  at phase 0, 0.25, 0.5 for hydrodynamic wind with $\dot M=4.0 \times 10^{-7} M_\odot {\rm yr}^{-1}$
  illuminated by a 10$^{36}$ erg s$^{-1}$ X-ray source.}
\end{figure*}

Figures \ref{fig13}  shows the spectra and polarization produced
by the hydrodynamic wind model with   $\dot M=4.0 \times 10^{-7} M_\odot {\rm yr}^{-1}$
at orbital phases 0, 0.25 and 0.5.  Comparison
with figure \ref{fig10} shows that the hydrodynamic model produces a spectrum which is 
 similar to that from 
a smooth spherical wind. However, the degree of polarization is significantly 
greater, due to the departure from spherical symmetry created by the X-ray 
source. In particular, the spectrum during eclipse is greater, owing to the 
fact that the wake structure extends beyond the disk of the primary star 
during eclipse. The spherical wind produces fractional polarization averaged 
over the 0.1 – 10 keV energy band, of at most 10$\%$ during eclipse. The spherical 
wind produces polarization values which are much less than this at mid-eclipse. 
The hydrodynamic wind produces fractional polarization of approximately 21$\%$ at mid-eclipse.

Polarization calculations for hydrodynamic models have been carried out for
the two mass loss rates shown in figure \ref{fig11}, and for several choices
of ionizing X-ray luminosity.  Note that the hydrodynamic models assume a fixed X-ray 
luminosity (used to calculate X-ray heating and ionization within the simulation) 
of 10$^{36}$ erg/s, independent of the X-ray luminosity used to calculate the 
spectrum and polarization.  Moreover, the prescribed X-ray luminosity is 
not generally consistent with the mass accretion rate derived from the hydrodynamic 
simulation, and thus these models are not fully self-consistent. Nonetheless, 
they serve to illustrate the behavior of the polarization and its dependence 
on wind density and X-ray luminosity.  

Values for polarization fractions during
eclipse are shown in table \ref{table1}.  These are averages of the linear polarization
over energy in the range from 0.1 -- 10 keV.  Polarization values reflect competing effects.
For highly ionized gas, the wind is essentially transparent and the polarization is due to
Thomson scattering.  The polarization fraction is proportional to the Thomson depth (when small)
and also to departures from circular symmetry of the scattering material in the plane of the sky.
Partially ionized gas can produce larger polarization, owing to the greater cross section associated
with resonance scattering, although each line has very limited spectral range and the ensemble of
lines for any given model seldom has a width which exceeds $\Delta \varepsilon/\varepsilon \sim 0.1$.  On the other hand,
partially ionized  and near-neutral gas also is affected by photoelectric absorption.  This limits
the size of the X-ray scattering region, and tends to reduce the net polarization.
Table \ref{table1} shows that, for the limited range of parameters spanned by our models,
more highly ionized models tend to produce greater polarization.  That is, the effects of photoelectric
absorption offset the effects of resonance scattering in partially ionized gas, leading to very weak
dependence of the polarization on X-ray luminosity when the luminosity is not large.
We expect that at very low X-ray luminosities this effect will be even stronger, since the
partially ionized zone containing the resonance scattering gas will shrink and become more round,
and photoelectric absorption will remove a larger fraction of photons.

\begin{figure*}[p] 
\includegraphics*[angle=90, scale=0.6]{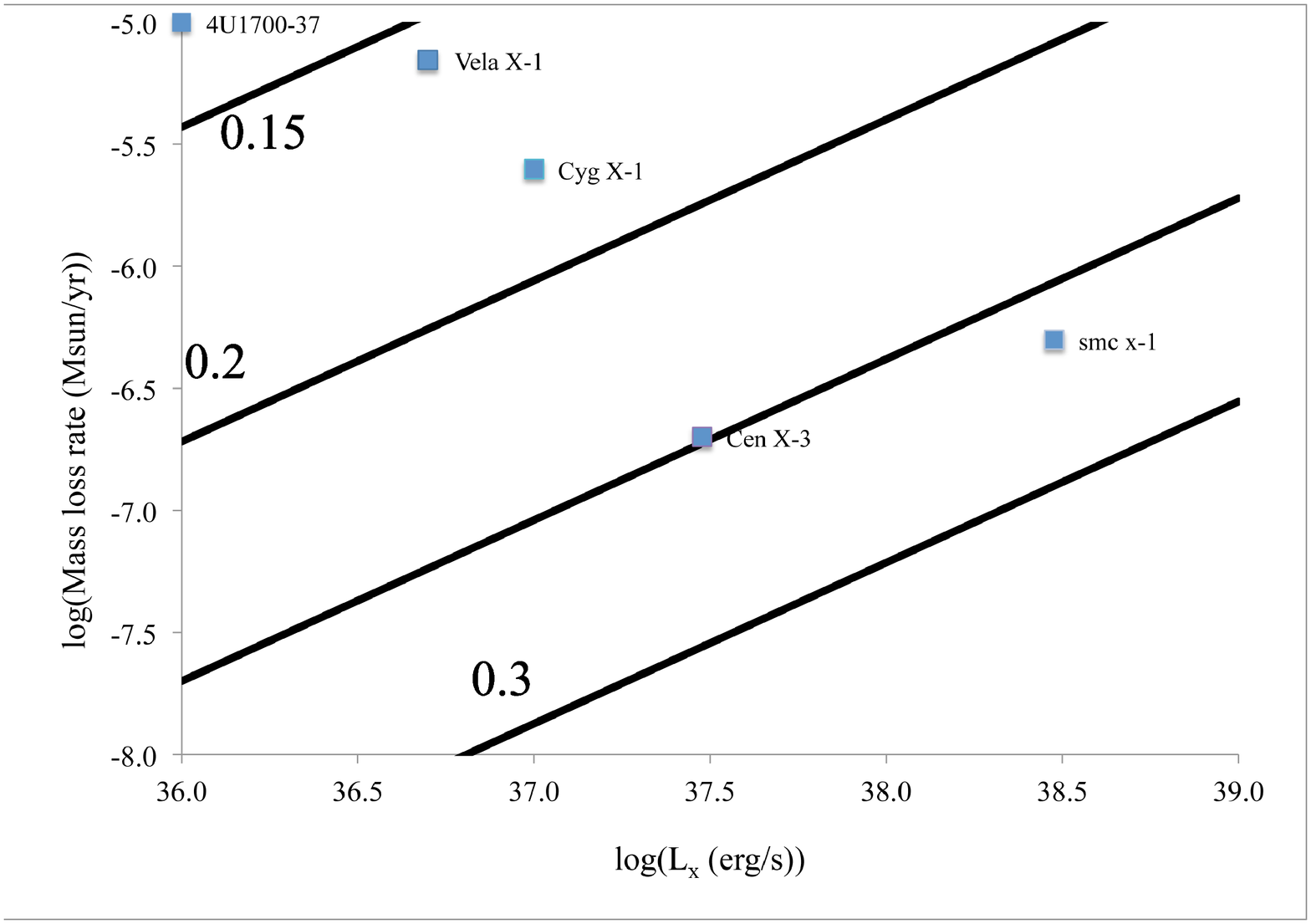}
\caption{\label{fig14} Diagram showing estimated polarization fractions in lines and continuum at conjunction vs. quadrature for
the objects in table \ref{table2}.}
\end{figure*}

\section{Discussion}
\label{sec5}

Our results so far on the polarization produced by wind scattering in HMXBs can be summarized as follows:
Polarization depends on inclination: high inclination produces variable polarization fraction plus
constant angle; low inclination produces constant polarization fraction plus variable polarization position angle.
The maximum attainable continuum polarization fraction scales approximately proportional to electron scattering optical depth, for
depths less than unity; this is partly a consequence of our single scattering assumption and must be suitably modified
if multiple scattering is important.
At high inclination, polarization fraction of the scattered radiation ($P_S$) at phase 0 and 0.5 is less than 
at phases 0.25 or 0.75; for a spherical wind the phase 0 and 0.5 polarization is zero.
Polarization fraction of the total radiation, including unscattered, is small for the parameters considered
here for all orbital phases out of eclipse.  A spherical wind thus produces narrow intervals of high polarization
during eclipse but away from phase 0.
Resonance line optical depths are greater than for electron scattering, and so can produce greater
linear polarization.  On the other hand, the optical depths are often large, and the
X-ray scattering regions tend to be small, i.e. nearly circular around the compact object, producing
smaller orbital phase modulation.  The hydrodynamics of the interaction between the wind and the compact object breaks
the spherical symmetry and increases the net polarization.  

Predictions of the polarization signatures for particular known HMXBs require
hydrodynamic simulations for each system incoporating known physical parameters:  the
sizes and masses of the components, the primary wind mass loss rate and the X-ray source
luminosity and spectrum.  These would then need to be analyzed using a transfer calculation
such as the single scattering calculations we have presented.
In this paper we have presented generic models for the wind densities and X-ray source properties,
both for spherical winds and incoporating the hydrodynamic effects of the
two gravitating centers.  These generic models can be used to infer, in a very simple way, the
polarization signatures expected from various sources.

Table \ref{table2} shows the known parameters of the 5 brightest and best-studied HMXBs.   The
relevant quantities are the source X-ray luminosity, primary type, mass loss rate, and average
X-ray luminosity.  These quantities are taken from \citet{Cont78} and from
\citet{Kape98}.  We do not include several systems which generally have smaller observed X-ray fluxes
or less well constrained properties.

With these quantities we can use our spherical wind models to
crudely predict the polarization fractions and orbital phase modulation of the polarization for these
sources.  We do this in the following way:  we use the results in table \ref{table1}
to construct an approximate scaling of mid-eclipse net polarization with the wind mass loss rate
and X-ray luminosity.   We then apply this to the known HMXBs in table 1.   The results
are shown graphically in figure \ref{fig14}.  The most relevant quantities for each of the table 1 HMXBs, X-ray
luminosity and wind mass-loss rate, are plotted on the axes and the values for each object are plotted
as points.  Contours of constant predicted mid-eclpise polarization are shown as solid curves, and labeled.  This
shows that high polarizations are expected for some objects, those with the strongest winds and weakest
X-rays generally.  This demonstrates that polarization fractions in the range
from 5 -- 30 $\%$ are expected at mid-eclipse for these systems.  More detailed information is contained in the
spectra and the time variation of the net polarization during eclipse; interpretation of these signals
requires modeling which is tailored to each particular system.

\acknowledgements  Support was provided through grant 10-ATP10-0171 through the NASA astrophysics theory program.

\begin{deluxetable}{crrr}
\tabletypesize{\scriptsize}
\tablecaption{Hydrodynamic Model Results.  Values for the polarization fraction are given for several hydrodynamic
  models described in the text. \label{table1}}
\tablewidth{0pt}
\tablehead{
\colhead{Model}&\colhead{$\dot M$}&\colhead{L$_{x}$}&\colhead{P$_{\phi=0.5}$} }
\startdata
  &$M_\odot {\rm yr}^{-1}$ & erg s$^{-1}$&\\
1& 4 $\times 10^{-7}$ & 10$^{38}$&0.25\\
2& 4 $\times 10^{-7}$ & 10$^{37}$&0.25\\
3& 4 $\times 10^{-7}$ & 10$^{39}$&0.29\\
4& 1.7 $\times 10^{-6}$ & 10$^{38}$&0.22\\
\enddata
\end{deluxetable}

\begin{deluxetable}{crrr}
\tabletypesize{\scriptsize}
\tablecaption{Sample HMXBs and Properties Needed for Predicting Wind Polarization.  \label{table2}}
\tablewidth{0pt}
\tablehead{
\colhead{object}&\colhead{Sp. type}&\colhead{$\dot M_{wind}$}&\colhead{L$_x$} }
\startdata
  &  &$M_\odot {\rm yr}^{-1}$&erg s$^{-1}$\\
Vela X-1&B0.5 Iab&7 $\times 10^{-6}$&0.05\\
Cen X-3&O6.5 II-III&2 $\times 10^{-7}$&0.3\\
Cyg X-1&O9.7 Iab&2.5 $\times 10^{-6}$&0.1\\
4U1700-37&O6.5 Iaf+&1 $\times 10^{-5}$&0.01\\
smc x-1&B0.6 Iab&5 $\times 10^{-7}$&3 \\
\enddata
\end{deluxetable}


\appendix
\begin{center}
      {\bf APPENDIX}
    \end{center}

Illustrative analytic results can be derived for an idealized HMXB system:  a spherical primary 
star with radius $R_{\*} \simeq 10^{12}$ cm in a circular orbit with an X-ray source with 
separation $a \simeq 1.5 R_{\*}$.   
We view the system from some phase angle $\Theta_V$ relative to the line of centers 
where $\Theta_V=0$ corresponds to superior conjunction of the X-ray source.  In addition, 
the system can have inclination $i$ defined such that $i=\pi/2$ corresponds to viewing in 
the orbital plane.  The X-ray source has a luminosity $L_X \simeq 10^{37}$ erg s$^{-1}$ and 
illuminates the primary star and wind isotropically.   The wind has a mass loss rate 
$\dot M \simeq 10^{-8} M_\odot$ yr$^{-1}$, a terminal velocity $v_\infty \simeq 10^8$cm s$^{-1}$,
and a velocity law $v(r)$ which depends only on the distance from the center of the star.
If so, the effects of scattering in the wind depend on the wind column density, which can be 
characterized by the column density $N=\int_{a}^{\infty} n(r)dr$ and an approximate 
value for this is $N=n_x a$ where $n_x=\frac{\dot M/\mu m_H}{4\pi a^2 v}$ is the density at
the X-ray source, $\mu$ is the mean particle weight, and $m_H$ is the hydrogen mass.  For the fiducial parameters given above, 
$n_x = 3.2 \times 10^8$ cm$^{-3}$ $\dot M_{8} v_8^{-1} a_{12}^{-2}$ and $N=3.2 \times 10^{20}$ cm$^{-2}$  
$\dot M_{8} v_8^{-1} a_{12}^{-1}$.  The ratio of the total intensity of scattered radiation to 
the total emitted radiation can be characterized by the quantity 
$L_{scatt}/L_X \sim N \sigma \simeq 2 \times 10^{-4} \sigma/\sigma_{Th} \dot M_{8} v_8^{-1} a_{12}^{-1}$
where  $\sigma$ is a characteristic scattering cross section and $\sigma_{Th}$ is the Thomson cross section.

We consider a coordinate system in which the observer is along the $\hat{\bf y}$ axis.  Then the scattering 
plane is perpendicular to the (x,z) plane.  
We can rewrite equation \ref{eq1} as:

\begin{equation}
\label{eq2}
\left\{
\begin{matrix}
L  \\
Q  \\
U 
\end{matrix}
\right\}
=L_0 \sigma \int dV \frac{n({\bf r})}{r_x^2} 
\left\{
\begin{matrix}
1 + \cos^2\chi  \\
\sin^2\chi \cos(2\gamma) \\ 
\sin^2\chi \sin(2\gamma)  
\end{matrix}
\right\}
\end{equation}

%

\noindent For the purposes of computation, a more convenient
expression is in terms of cylindrical coordinates centered on the X-ray source, with the 
observer on the axis at infinity:

\begin{equation}
\label{eq4}
\left\{
\begin{matrix}
L  \\
Q  \\
U 
\end{matrix}
\right\}
=L_0 \sigma \int_0^{2\pi} d\gamma \int_0^{\infty} p dp \int_{-\infty}^{\infty} d\zeta \frac{n({\bf r})}{(p^2+\zeta^2)^2} 
\left\{
\begin{matrix}
(p^2+2\zeta^2)  \\
p^2 cos(2\gamma) \\ 
p^3 sin(2\gamma)  
\end{matrix}
\right\}
\end{equation}

\noindent where $(p,\gamma,\zeta)$ are the radial, angular and axial cylindrical coordinates with the axis along y.
These are indicated on figure \ref{fig1}.  In the case 
a star located at a distance $a$ from the X-ray source, orbiting in a plane with a normal which is inclined at an 
agle $i$ with respect to line of sight, and with an orbital phase described by an angle $\theta_v$, this 
expression can be rewritten in terms of coordinates centered on the star:

\begin{equation}
p=r((x \cos\theta_v+(y + \alpha)\sin\theta_v)^2
+ (z \sin i +  (y+ \alpha)\cos\theta_v 
    - x \sin\theta_v)\cos i)^2)^{1/2}
\end{equation}

\begin{equation}
\gamma={\rm tan}^{-1}\left(\frac{z \sin i + (y +  \alpha)\cos\theta_v
  - x \sin\theta_v)\cos i}
{x \cos\theta_v+(y + \alpha)\sin\theta_v}\right)
\end{equation}

\begin{equation}
\zeta= - r ((y+ \alpha)\cos\theta_v
  - x \sin\theta_v)\sin i
+ z \cos i
\end{equation}

%
%

\noindent where $x, y, z$ are cartesian coordinates  centered on the star $r$
is the distance from the center of the star and $\alpha=a/r$.
This can be evaluated analytically for special cases.

In the case of a single star, $a$=0, $\theta_v$=0, and if the density distribution is axially symmetric, 
\citet{Brow77} have shown that

\begin{equation}
\left\{
\begin{matrix}
L  \\
Q  \\
U 
\end{matrix}
\right\}
=L_0 \sigma \int_0^{\infty}  dr \int_{-1}^{1} \sin\theta d\theta n(r,\theta)
\left\{
\begin{matrix}
2 (1+\cos^2\theta)+\sin^2i (1-3\cos^2\theta)\\
\sin^2i (1-3\cos^2\theta) \\ 
0
\end{matrix}
\right\}
\end{equation}

\noindent so that the fraction varies from zero for zero inclination to a maximum value 
determined by the departure of the density distribution from spherical. 

From equation \ref{eq4} it is apparent that,  in the case where the 
density is independent of $\gamma$, the net polarization is zero and the surfaces of constant 
polarization in the integrand are circular.  
This is expected to be the case for a binary system where the gas density is circularly symmetric 
around the line of centers, and when viewed at conjunction from the orbital plane (i.e. $\theta_v$=0 or $\pi$/2 
and $i$=$\pi/2$).   In the case of a binary system viewed along the angular momentum axis, 
i.e. $i$=0, the intensities can be shown to be:

\begin{equation}
\left\{
\begin{matrix}
L  \\
Q  \\
U 
\end{matrix}
\right\}
=
\left\{
\begin{matrix}
\mathcal{I}_0 \\
\mathcal{I}_1 \cos2\theta_v+{\mathcal{I}_2 \sin}2\theta_v\\
\mathcal{I}_1 \sin2\theta_v+{\mathcal{I}_2 \cos}2\theta_v
\end{matrix}
\right\}
\end{equation}

\noindent where

\begin{equation}
\mathcal{I}_0=L_0 \sigma \int_0^{\infty}r^2  dr \int_{0}^{\pi} \sin\theta d\theta n(r,\theta, \phi)
\frac{1}{(r^2 + 2 a r \sin\theta\sin^2\phi + a^2)}
\end{equation}

\begin{equation}
\mathcal{I}_1=L_0 \sigma \int_0^{\infty}r^2  dr \int_{0}^{\pi} \sin\theta d\theta n(r,\theta, \phi)
\frac{(r^2 \sin^2\theta \cos^2\phi - (r \sin\theta \sin\phi + a)^2)}
{(r^2 + 2 a r \sin\theta\sin^2\phi + a^2)^2}
\end{equation}

\begin{equation}
\mathcal{I}_2=L_0 \sigma \int_0^{\infty}r^2  dr \int_{0}^{\pi} \sin\theta d\theta n(r,\theta, \phi)
\frac{r^2 \sin^2\theta \sin\phi \cos\phi}
{(r^2 + 2 a r \sin\theta\sin^2\phi + a^2)^2}
\end{equation}

\noindent and $\mathcal{I}_2=0$ by symmetry if the density distribution is independent of $\phi$.
This clearly represents a circle in the $(Q,U)$ plane.  This, plus more general cases, were 
explored by \citet{Brow78}.

In the case of a binary viewed from within the orbital plane, it is straightforward to show
that the polarization is zero at phases 0 and 0.5, i.e. superior and inferior conjunction of the
X-ray source.  At phase 0.25 or 0.75, i.e. quadrature, the stokes quantities can be written:

\begin{equation}
\label{eq6}
\left\{
\begin{matrix}
L  \\
Q  \\
U 
\end{matrix}
\right\}
=\frac{L_0 \sigma \pi}{8a} \int_0^{\infty} r dr n(r)
\left\{
\begin{matrix}
(18r^2+15a^2-\frac{r^4}{a^2})\frac{4ar}{(r^2-a^2)^2}-\frac{4r}{a}+(\frac{2r^2}{a^2}-6){\rm ln}\left(\frac{r+a}{r-a}\right))  \\
(22r^2+19a^2+\frac{9r^4}{a^2})\frac{4ar}{(r^2-a^2)^2}+\frac{36r}{a}-(\frac{18r^2}{a^2}+26){\rm ln}\left(\frac{r+a}{r-a}\right)) \\ 
0
\end{matrix}
\right\}
\end{equation}

\noindent This can be evaluated analytically in the limit that the density distribution is a thin shell at $r=r_*$ in which
case the net polarization $Q/L$ varies from unity for $a>>r_*$ to 0.886 for  $a=r_*$

%
%

%
%
%
%
%

%
%

\bibliographystyle{apj}

\end{document}